\documentclass[10pt,a4paper]{article}
\usepackage{graphicx}
\usepackage{amsmath,amssymb}
\usepackage{amsfonts}
\begin{document}
\textwidth=135mm
 \textheight=200mm
\begin{center}
{\bfseries Self - Consistent Description of $e^- e^+ \gamma$- Plasma Created
from the Vacuum in a Strong Electric Laser Field}
\vskip 5mm
D.B.~Blaschke$^{a,b}$, S.M.~Schmidt$^{c,d}$, S.A.~Smolyansky$^{e,}\footnote{{\small E-mail: smol@sgu.ru}}$ and A.V. Tarakanov$^e$
\vskip 5mm
$^a${\small {\it University of Wroc{\l}aw, 50-204 Wroc{\l}aw, Poland}} \\
$^b${\small {\it Joint Institute for Nuclear Research, 141980 Dubna, Russia}}\\
$^c${\small {\it Forschungszentrum J\"ulich GmbH, D-52428 J\"ulich, Germany}}\\
$^d${\small {\it Technische Universit\"at Dortmund, D-44221 Dortmund, Germany}}
\\
$^e${\small {\it Physics Department of Saratov State University,
410026, Saratov, Russia}}\\
\end{center}
\vskip 5mm
\centerline{\bf Abstract}
In the present work a closed system of kinetic equations is obtained for the
description of the vacuum creation of an electron - positron plasma and
secondary photons due to a strong laser field.
An estimate for the photon energy distribution is obtained.
In the Markovian approximation the photon distribution has a 1/k spectrum
(flicker noise).
\vskip 10mm
\subsection*{\label{sec:intro}Introduction}


Up to now, the Schwinger effect \cite{SHS} has resisted an experimental
verification. This is basically due to the huge critical field strength
$E_c$ which could not yet be reached in the laboratory.
Recently, main attention was devoted to theoretical studies of
pair creation by time-varying electric fields \cite{Brezin,PNN,Marinov}
where sufficiently strong electric fields can be achieved at modern
high-intensity laser facilities.
It has been shown \cite{Brezin,PNN,Marinov,Popov} that pair
creation by a single laser pulse with $E\ll E_c$ could hardly be observed.
More optimistic results have been obtained for a X-ray free electron lasers
\cite{Ringwald,V01,V02} and for counter-propagating beams of optical lasers
\cite{Av02,Tar,20}.
It is obvious, that for subcritical fields $E\ll E_{c}$ the electron - positron
excitations have quasiparticle character and S - matrix methods can not be
applied \cite{SarRev} and existing estimates \cite{V02} are not reliable.
An adequate method is the kinetic theory.

We will construct here the system of kinetic equations for a selfconsistent
description of the electron - positron - photon system generated from the
vacuum by a time-dependent electric field.
As a first step, we will consider the one-photon annihilation process only,
which in the presence of a strong field is not forbidden \cite{Rit}.

Below we will assume an external electric field
$A^{\mu}(t)=(0,\mathbf{A}_{\rm ext}(t))$ that is spatially homogeneous.
In the case of a strong external electric field $\mathbf{A}_{\rm ext}(t)$ also
some internal field $\mathbf{A}_{\rm int}(t)$ will be generated.
The total acting field will be equal to
$\mathbf{A}(t)=\mathbf{A}_{\rm int}(t)+ \mathbf{A}_{\rm ext}(t)$
and this field is quasiclassical.
Some fluctuations of the electromagnetic field can arise against this
background.
They can be interpreted as photon excitations.
These photons, in principle, can be registered far from the active zone.

\subsection*{\label{sec:EPS}1 Electron - positron sector}
In general, the complete system of equations for a selfconsistent description
of the electron - positron - photon plasma consist of:
1) the KEs for the electron and positron quasiparticle components with
distribution functions $f_{e,p}(\mathbf{p},t)$ in the presence of a total
electric field $\mathbf{E}=-\dot{\mathbf{A}}(t)$,
2) the KE for the photon component, and
3) the Maxwell equation for the internal field $\mathbf{A}_{\rm int}(t)$.
We assume the electroneutrality condition
$f_{e}(\mathbf{p},t)=f^{c}(-\mathbf{p},t)=f(\mathbf{p},t)$
to be fulfilled.

We start from the standard QED Lagrangian
$\mathcal{L}=\mathcal{L}_{0}+\mathcal{L}^{\prime}$
taking into account in $\mathcal{L}_0$ the interaction with a quasiclassical
($A_\mu(t)$) and in $\mathcal{L}^{\prime}$ with a quantized ($\hat{A}_\mu(t)$)
electromagnetic field,
\begin{equation}\label{2e}
\mathcal{L}_0 = \frac{i}{2} \{ \overline{\psi} \gamma^\mu D_{\mu}
\Psi - (D^{*}_{\mu} \overline{\psi}) \gamma^\mu \Psi\} -m
\overline{\psi} \psi \quad ,\quad
\mathcal{L}^{\prime} = -e\overline{\psi}\gamma^\mu \hat{A}_{\mu}\Psi,
\end{equation}
where $D_{\mu}=\partial_{\mu} + ie{A}_{\mu}(t)$.
It is assumed that the intensity of the quantized field is rather weak so that
there is no backreaction influence on the state of the system.
In other words, the electron - positron system plays the role of a photon
source only.
The in - vacuum $|in\rangle=|\rangle$ is defined such that
$\langle\hat{A}_{\mu}(x)\rangle=0$.
Below we will not consider the backreaction problem because for subcritical
fields $E\ll E_{c}$ the internal field is negligible.

The kinetics of electron - positron vacuum pair creation due to a linearly
polarized electric field was studied in a large number of works.
See, e.g., Refs. \cite{Pervushin,Vin01} and the works quoted therein.
The corresponding generalization to the case of an arbitrarily polarized time
dependent electric field was obtained in \cite{Kiel04,Fil08,PS}.
The oscillator representation in its different realizations
\cite{Pervushin,Vin01} leads to the nonstationary orthonormalized spinor basis
\begin{align}\label{5e}
u^+_1(\mathbf{p},t) = A(\mathbf{p}) \begin{bmatrix} \omega_+ , 0 , p^3
,p_- \end{bmatrix}, && u^+_2(\mathbf{p},t) = A(\mathbf{p})
\begin{bmatrix} 0, \omega_+ ,  p_+
, -p^3 \end{bmatrix}, \nonumber \\
v^+_1(-\mathbf{p},t) = A(\mathbf{p}) \begin{bmatrix}  -p^3, - p_- ,
\omega_+ , 0 \end{bmatrix}, && v^+_2(-\mathbf{p},t) = A(\mathbf{p})
\begin{bmatrix} -p_+ , p^3 ,0, \omega_+
\end{bmatrix},
\end{align}
where $\omega(\mathbf{p},t)=\sqrt{m^2 +\mathbf{P}^2}=\omega$,
$\mathbf{P}=\mathbf{p}-e\mathbf{A}$,
$p_{\pm} = p^1 \pm i p^2$,
$\omega_+ =\omega +m$ and
$A(\mathbf{p}) = [2\omega\omega_+]^{-1/2}$.

Finally, the Dirac equation in the presence of an external quasiparticle field
$\mathbf{A}_{\rm ext}(t)$ generates the Heisenberg-like equations of motion
for the construction operators
\begin{align} \label{9e}
\dot{a}(\mathbf{p},t) &=
- U_{(1)}(\mathbf{p},t)a(\mathbf{p},t)
- U_{(2)}(\mathbf{p},t)b^+(-\mathbf{p},t)
- i\omega(\mathbf{p},t)a(\mathbf{p},t)~, \nonumber\\
\dot{b}(-\mathbf{p},t) &=
  b(-\mathbf{p},t) U_{(1)}(\mathbf{p},t)
+ a^+(\mathbf{p},t)U_{(2)}(\mathbf{p},t)
- i\omega (\mathbf{p},t)b(-\mathbf{p},t)~,
\end{align}
where in the representation (\ref{5e}) the matrices are
$
U_{(1)}(\mathbf{p},t)  =
i \omega a [\mathbf{p}\mathbf{E}]\mathbf{{\boldsymbol\sigma}}
=iU_{k}\sigma_{k}$ and
$
U_{(2)}(\mathbf{p},t) =
a [\mathbf{P}(\mathbf{P}\mathbf{E})-\mathbf{E}\omega\omega_+]
\mathbf{\boldsymbol\sigma}
$
with
$ a=e/(2\omega^2\omega_+) $.

KEs for the electron - positron component of the plasma follow from the
equations of motion (\ref{9e}) and the definitions of the electron and positron
distribution functions in the instantaneous representation
\begin{equation}
\label{11e}
f_{\alpha\beta}(\mathbf{p},t) =
\langle a^+_{\beta}(\mathbf{p},t) a_{\alpha}(\mathbf{p},t)\rangle ~,~~
{f}^c_{\alpha\beta}(\mathbf{p},t) =
\langle b_{\beta}(-\mathbf{p},t) b^+_{\alpha}(-\mathbf{p},t)\rangle
\end{equation}
and also the two additional functions
\begin{equation}\label{12e}
f^{(+)}_{\alpha\beta}(\mathbf{p},t)  =
\langle a^{+}_{\beta}(\mathbf{p},t) b^{+}_{\alpha}(-\mathbf{p},t)\rangle ~,~~
f_{\alpha\beta}^{(-)}(\mathbf{p},t)  =
\langle b_{\beta}(-\mathbf{p},t)a_{\alpha}(\mathbf{p},t)\rangle
\end{equation}
describing vacuum polarization.
The system of KEs is then \cite{Kiel04,Fil08,PS}
\begin{align}\nonumber
\dot{f} &= [f,U_{(1)}] - \bigl( U_{(2)} f^{(+)} + f^{(-)}U_{(2)}
\bigr) ,\\ \label{13e}
 \dot{f}^c &= [f^c,U_{(1)}] + \bigl(
f^{(+)}U_{(2)} + U_{(2)} f^{(-)}\bigr),\\
\dot{f}^{(+)}& =  [ {f}^{(+)} , U_{(1)} ] + \bigl( U_{(2)} f - f^c
U_{(2)}
\bigr) + 2i\omega f^{(+)}, \nonumber \\
\dot{f}^{(-)} &= [ {f}^{(-)} , U_{(1)} ] + \bigl( f U_{(2)} -
U_{(2)} f^c \bigr) - 2i\omega f^{(-)}. \nonumber
\end{align}
If the standard decomposition in the basis of Pauli matrices is used
$ f = f_0+f_k \sigma_k$, where $ f_0 = {\rm Tr}\{f\}/2$, and
$f_k={\rm Tr} \{f\sigma_k\}/2$,
the system  KEs (\ref{13e}) can be rewritten in the spin representation.
Some of the simplest applications of the corresponding system of KEs can be
found in \cite{Kiel04,Fil08,PS}.

\subsection*{\label{sec:PHS}2 Photon sector}
In order to construct the photon kinetics it is necessary to
derive the corresponding generalization of the quasiparticle formalism
developed in Sect. 1, 
using it as a nonperturbative basis.
The interaction with the quantized electromagnetic field $\hat{A}_\mu (x)$ in
the fermion sector of the theory is introduced by means of the substitution
$H_0 \rightarrow H_0 +H^{\prime}$ in the Heisenberg-like equation of motion
(\ref{9e}),
\begin{equation}\label{16e}
\dot{a}(\mathbf{p},t) + U_{(1)}(\mathbf{p},t)a(\mathbf{p},t) +
U_{(2)}(\mathbf{p},t)b^+(-\mathbf{p},t) = -i [a(\mathbf{p},t),H_0+H^{\prime}],
\end{equation}
where $H_0$ is the Hamiltonian of the fermion field in the quasiparticle
representation and $H^{\prime}$ is the usual Hamiltonian of interaction with
the quantized field.
The time dependence of $H^{\prime}(t)$ makes manifest the nonstationarity of
the system that is also reflected in the  decomposition of the field operators
$\Psi,\,\Psi^{+}$ in the nonstationary basis (\ref{5e}).
The same source (external field) induces the nonstationarity of the
quantized electromagnetic field.
However, this does not alter the  mass shell of the photon field, $k^2 =0$
(in contrast to electron - positron field, $\omega(\mathbf{p},t)$), so that
the standard decomposition is valid,
\begin{equation}\label{18e}
    \hat{A}_{\mu}(x)=(2\pi)^{-3/2}\int\frac{d^{3}k}{\sqrt{2k}}
	A_{\mu}(\mathbf{k},t)e^{-i\mathbf{k}\mathbf{x}},
\end{equation}
where
$
A_{\mu}(\mathbf{k},t)=A_{\mu}^{(+)}(\mathbf{k},t)+A_{\mu}^{(-)}(-\mathbf{k},t).
$

The system of Heisenberg-like equations of motion taking into account the
photon subsystem can be written now in explicit form. For example,
\begin{eqnarray}
      \label{24e}
     & iA^{(\pm)}_r (\mathbf{k},t) =
\mp k A^{(\pm)}_r (\mathbf{k},t)\mp e(2\pi)^{-3/2}\frac{1}{\sqrt{2k}}
\int d^3 p_1 d^3 p_2\; \delta(\mathbf{p}_1 -\mathbf{p}_2 \mp\mathbf{k})
\nonumber\\
   &\Bigl\{ a^+ (\mathbf{p}_1,t)[\bar{u}u]^{r}(\mathbf{p}_1 ,\mathbf{p}_2 ,t)
a(\mathbf{p}_2 ,t)+
a^+ (\mathbf{p}_1,t)[\bar{u}v]^{r}(\mathbf{p}_1 ,\mathbf{p}_2 ,t)
b^+ (-\mathbf{p}_2 ,t)
   \nonumber\\
   &+  b(-\mathbf{p}_1,t)[\bar{v}u]^{r}(\mathbf{p}_1 ,\mathbf{p}_2 ,t)
a(\mathbf{p}_2 ,t)
+ b(-\mathbf{p}_1,t)[\bar{v}v]^{r}(\mathbf{p}_1 ,\mathbf{p}_2 ,t)
b^+ (-\mathbf{p}_2 ,t)\Bigr\}.
 \end{eqnarray}
Here and below the vectors $\mathbf{p}_1 ,\mathbf{p}_2 , \ldots$ denote the
canonical momenta of fermions and $\mathbf{k}_1 ,\mathbf{k}_2 ,\ldots$
correspond to the momenta of photons;\\
$
[\bar{\xi}\eta]^{r}_{\beta\alpha}(\mathbf{p}_1 ,\mathbf{p}_2 ;t)
=\bar{\xi}_{\alpha} (\mathbf{p}_1,t)\gamma^{\mu}
\eta_{\beta}(\mathbf{p}_2,t)e^r _\mu,
$
and
$\vec{e}^{\;1} ,\vec{e}^{\;2}$ are the polarization unit vectors,
$\vec{e}^{\;3} =\vec{k}/k$ and $e^0 _\mu =\delta^0 _\mu$.
The photon correlation function is defined as
\begin{equation}\label{25e}
    F_{rr'}(\mathbf{k},\mathbf{k}',t)=
\langle A^{+}_{r}(\mathbf{k},t)A^{-}_{r'}(\mathbf{k}',t)\rangle.
\end{equation}
With help Eq. (\ref{24e}) we can then obtain the first equation of the BBGKY
hierarchy
\begin{eqnarray}
\dot{F}_{rr'}(\mathbf{k},\mathbf{k}',t) &=&
ie(2\pi)^{-3/2}\sum_{\alpha\beta}\int d^3 p_1 d^3 p_2
\Big\{-\frac{1}{\sqrt{2k}} \delta(\mathbf{p}_1 -\mathbf{p}_2 -\mathbf{k})
\nonumber  \\
   &\times&\bigg[[\bar{u}u]^{r}_{\beta\alpha}(\mathbf{p}_1 ,\mathbf{p}_2 ,t)
\langle a^{+}_{\alpha}(\mathbf{p}_{1},t)a_{\beta}(\mathbf{p}_2 ,t)
        A_r ^{(-)} (\mathbf{k},t)\rangle
\nonumber \\
   &+& [\bar{u}v]^{r}_{\beta\alpha}(\mathbf{p}_1 ,\mathbf{p}_2 ,t)
\langle a^{+}_{\alpha}(\mathbf{p}_{1},t)b_{\beta}^+ (-\mathbf{p}_2 ,t)
        A_r ^{(-)} (\mathbf{k},t)\rangle
\nonumber \\
   &+& [\bar{v}u]^{r}_{\beta\alpha}(\mathbf{p}_1 ,\mathbf{p}_2 ,t)
\langle b_{\alpha}(-\mathbf{p}_{1},t)a_{\beta}(\mathbf{p}_2 ,t)
        A_r ^{(-)} (\mathbf{k},t)\rangle
\nonumber \\
   &+& [\bar{v}v]^{r}_{\beta\alpha}(\mathbf{p}_1 ,\mathbf{p}_2 ,t)
\langle b_{\alpha}(-\mathbf{p}_{1},t)b_{\beta}^+ (-\mathbf{p}_2 ,t)
        A_r ^{(-)} (\mathbf{k},t)\rangle \bigg]
\nonumber \\
  &+& \frac{1}{\sqrt{2k}} \delta(\mathbf{p}_1 -\mathbf{p}_2 +\mathbf{k}')
\nonumber\\
 &\times&\bigg[[\bar{u}u]^{r'}_{\beta\alpha}(\mathbf{p}_1 ,\mathbf{p}_2 ,t)
\langle a^{+}_{\alpha}(\mathbf{p}_{1},t)a_{\beta}(\mathbf{p}_2 ,t)
        A_{r'} ^{(+)} (\mathbf{k}',t)\rangle
\nonumber \\
   &+& [\bar{u}v]^{r'}_{\beta\alpha}(\mathbf{p}_1 ,\mathbf{p}_2 ,t)
  \langle a^{+}_{\alpha}(\mathbf{p}_{1},t)b_{\beta}^+ (-\mathbf{p}_2 ,t)
	A_{r'} ^{(+)} (\mathbf{k}',t)\rangle
\nonumber \\
   &+& [\bar{v}u]^{r'}_{\beta\alpha}(\mathbf{p}_1 ,\mathbf{p}_2 ,t)
\langle b_{\alpha}(-\mathbf{p}_{1},t)a_{\beta} (\mathbf{p}_2 ,t)
	A_{r'} ^{(+)} (\mathbf{k}',t)\rangle
   \nonumber \\
   &+& [\bar{v}v]^{r'}_{\beta\alpha}(\mathbf{p}_1 ,\mathbf{p}_2 ,t)
\langle b_{\alpha}(-\mathbf{p}_{1},t)b_{\beta}^+ (-\mathbf{p}_2 ,t)
	A_{r'} ^{(+)} (\mathbf{k}',t)\rangle \bigg]\Big\}.
\label{26e}
\end{eqnarray}

\subsection*{3 Truncation procedure}
The simplest decoupling of the hierarchy,
\begin{equation}\label{27e}
    \langle a^{+}_{\alpha}(\mathbf{p}_{1},t)a_{\beta}(\mathbf{p}_2 ,t)
	A_r ^{(\pm)} (\mathbf{k},t)\rangle
\simeq\langle a^+_\alpha(\mathbf{p}_{1},t)a_\beta(\mathbf{p}_2,t)\rangle
	\langle A_r ^{(\pm)} (\mathbf{k},t)\rangle =0,
\end{equation}
is not effective.
Therefore, we will consider the equation at the second order for the
annihilation process (the next to last line in Eq.~(\ref{26e}))
and the reverse one (the fourth line).
We will not write the equations of the second level for these correlators
completely in view of their awkwardness and will discuss only the simplest
truncation scheme resulting in correlators of the type
\begin{eqnarray}
\langle a^+ (\mathbf{p}_{1},t)a(\mathbf{p}_2 ,t)
	A_r^{(\pm)} (\mathbf{k},t)A_{r'}^{(\pm)} (\mathbf{k}',t)\rangle
\simeq\nonumber\\
\langle a^+ (\mathbf{p}_{1},t)a(\mathbf{p}_2,t)\rangle
\langle A_r^{(\pm)}(\mathbf{k},t)A_{r'}^{(\pm)} (\mathbf{k}',t)\rangle,
\label{29e}
\end{eqnarray}
which occur in the RPA approximation as well.
We ignore  other processes as, e.g., the vacuum polarization effects
contained in correlators of the type
$ \langle a(\mathbf{p}_{1},t)b(-\mathbf{p}_2 ,t)
A_r ^{(\pm)} (\mathbf{k},t)A_{r'} ^{(\pm)} (\mathbf{k}',t)\rangle$
leading to the polarization functions (\ref{12e}).
Ignoring spin effects (see Sect.~2), 
the approximation (\ref{29e}) in combination with the diagonalization of the
photon and fermion correlation functions
\begin{eqnarray}\nonumber
    \langle A_{r}^{(+)}(\mathbf{k},t)A_{r'}^{(-)}(\mathbf{k}',t)\rangle =
\delta_{r r'}\delta(\mathbf{k}-\mathbf{k}')F_r (\mathbf{k},t),\\
    \label{31e}
    \langle a_{\alpha}^{(+)}(\mathbf{p},t)a_{\beta}(\mathbf{p}\, ',t)\rangle =
\delta_{\alpha \beta}\delta(\mathbf{p}-\mathbf{p}\, ')f(\mathbf{p},t)
\end{eqnarray}
leads to the following photon KE for zero initial condition
\begin{eqnarray}
  & \dot{F}(\mathbf{k},t) =-\frac{e^2}{2(2\pi)^3 k}\int d^3 p \int_{t_0}^{t}dt' K(\mathbf{p},\mathbf{p}-\mathbf{k};t,t')
   \bigl [1+F(\mathbf{k},t')\bigr ]\cdot\\
   \nonumber &\bigl [f(\mathbf{p},t')+f(\mathbf{p}-\mathbf{k},t')-1\bigr ]
   \cos \Bigl\{\int_{t'}^{t}d\tau \bigl [\omega(\mathbf{p},\tau)+\omega(\mathbf{p}-\mathbf{k},\tau)-k\bigr ]\Bigr\}, \label{32e}
    \end{eqnarray}
where the  nonlocal kernel is
    $
K(\mathbf{p},\mathbf{p}\, ';t,t')=
[\bar{v}u]^r _{\beta\alpha}(\mathbf{p},\mathbf{p}\, ';t)
[\bar{u}v]^r _{\alpha\beta}(\mathbf{p}\, ',\mathbf{p};t').
    $
The next step is based on the Markovian approximation that allows to ignore
the memory effect in the photon distribution, i.e. on r.h.s. of Eq.~(\ref{32e})
we replace $F(\mathbf{k},t')\rightarrow F(\mathbf{k},t)$.
That brings us to the following quadrature formula
\begin{equation}\label{34e}
   F(\mathbf{k},t)=\exp \left[\Phi (\mathbf{k},t)\right] -1
                   \simeq \Phi (\mathbf{k},t)~,
\end{equation}
where the last approximation is valid for the subcritical fields and
\begin{eqnarray}
 \nonumber
  &\Phi(\mathbf{k},t) =-\frac{e^2}{2(2\pi)^3 k}\int_{t_0}^{t}dt'
\int_{t_0}^{t'}dt''\int d^3 p K(\mathbf{p},\mathbf{p}-\mathbf{k};t',t'')\\
   \nonumber& \bigl [f(\mathbf{p},t'')+f(\mathbf{p}-\mathbf{k},t'')-1\bigr ]
    \cos \Bigl\{\int_{t''}^{t'}d\tau \bigl [\omega(\mathbf{p},\tau)
    +\omega(\mathbf{p}-\mathbf{k},\tau)-k\bigr ]\Bigr\}.
\label{35e}
    \end{eqnarray}
The kernel $K(\mathbf{p},\mathbf{p}\, ';t,t')$ is a slowly varying function of
the momentum arguments $\mathbf{p},\mathbf{p}\, '$ at fixed $t$ and $t'$.
There are also some complicated fast temporal oscillations on this background.
For a rough estimate of the effect let us substitute the kernel $K$ by its
average value
    $
       K(\mathbf{p},\mathbf{p}-\mathbf{k};t,t')\rightarrow K_0 =-5.
    $
In addition, we neglect the non-Markovian effect in the fermion distribution
functions.
Since the main part of fermions is created from vacuum with small momenta they
can be neglected due to the momentum and field ($E\ll E_c $) dependence in the
high frequency factor on the r.h.s. of Eq.~(\ref{35e}).
We obtain the result
    \begin{equation}\label{37e}
        F(\mathbf{k},t)=\frac{5e^2 n(t)}{2k \delta^2},
    \end{equation}
where $\delta=2 m-k$  is the frequency mismatch and
$n(t)=2\int d^3 pf(\mathbf{p},t)/(2\pi)^3$
is the pair density with the factor 2 from the spin degeneracy.
Thus, in the optical region $k\ll m$ the distribution (\ref{37e}) gives
$F(k)\sim 1/k$, which is characteristic for the flicker noise
(see, e.g., \cite{Kus}).
In the high frequency region $k\sim m$ the Markovian approximation is not
justified. Here, a more detailed investigation is necessary.

\subsection*{Summary}
Our main result in Eq.~(\ref{37e}) defines the frequency dependence of the
photon distribution by the factor $1/k$.
That this is a multi-photon process can be seen when one identifies the
frequency mismatch with the energy $Nk$ of the photon system which is
necessary for the energy conservation in the one-photon annihilation process:
we obtain $N\sim 2m/k$.
For optical lasers this is a huge number and therefore such kind of events 
are very rare.
This conclusion about the role of multi-photon processes is conform with the
analysis of the absorption coefficient of the electron - positron plasma
created from the vacuum in the infrared region \cite{BIP}.

\subsubsection*{Acknowledgements}
DBB and SAS acknowledge the financial support and hospitality of the
Research Centre Juelich.
The work of DBB was supported in part by the MNiSW grant No. N N 202 231837
and by RFFI grant No. 08-02-01003-a.

\end{document}